\documentclass[12pt, a4paper]{article}
\baselineskip=\normalbaselineskip

\usepackage{mathrsfs,amsbsy,amssymb,latexsym,amsfonts,amsmath,amsthm}
\usepackage{graphicx,color}
\usepackage[nosort]{cite}

\usepackage{bm}

\newcommand{\bbC}{{\mathbb{C}}}
\newcommand{\bbR}{{\mathbb{R}}}

\makeatletter
\catcode`\@=11
\@addtoreset{equation}{section}

\def\@seccntformat#1{\csname the#1\endcsname.~~}
\makeatother

\begin{document}

\begin{titlepage} 

\renewcommand{\thefootnote}{\fnsymbol{footnote}}
\begin{flushright}
%   KUNS-2885, RBRC-****, YITP-21-75
  KUNS-2885, YITP-21-75
\end{flushright}
\vspace*{1.0cm}

\begin{center}
{\Large \bf
Statistical analysis method 
for the worldvolume hybrid Monte Carlo algorithm 
}
\vspace{1.0cm}

\centerline{
{Masafumi Fukuma${}^1$}%
\footnote{
  E-mail address: fukuma@gauge.scphys.kyoto-u.ac.jp
}, 
{Nobuyuki Matsumoto${}^2$}%
\footnote{
  E-mail address: nobuyuki.matsumoto@riken.jp} 
and 
{Yusuke Namekawa${}^3$}%
\footnote{
  E-mail address: namekawa@yukawa.kyoto-u.ac.jp}
}

\vskip 0.8cm
  ${}^1${\it Department of Physics, Kyoto University,
  Kyoto 606-8502, Japan}
\vskip 0.1cm
  ${}^2${\it RIKEN/BNL Research center, Brookhaven National Laboratory,
  Upton, NY 11973, USA}
\vskip 0.1cm
  ${}^3${\it Yukawa Institute for Theoretical Physics,
  Kyoto University, 
  Kyoto 606-8502, Japan}
\vskip 1.2cm

\end{center}

%%%%%%%%%%%%%%%%%%%%%%%%%%%%%%%%%%%%%%%
\begin{abstract}
%%%%%%%%%%%%%%%%%%%%%%%%%%%%%%%%%%%%%%%

We discuss the statistical analysis method 
for the worldvolume hybrid Monte Carlo (WV-HMC) algorithm 
[arXiv:2012.08468], 
which was recently introduced 
to substantially reduce the computational cost 
of the tempered Lefschetz thimble method. 
In the WV-HMC algorithm, 
the configuration space is a continuous accumulation (worldvolume) 
of deformed integration surfaces, 
and sample averages are considered 
for various subregions in the worldvolume. 
We prove that, 
if a sample in the worldvolume is generated as a Markov chain, 
then the subsample in the subregion can also be regarded as a Markov chain. 
This ensures the application of the standard statistical techniques 
to the WV-HMC algorithm. 
We particularly investigate the autocorrelation times 
for the Markov chains in various subregions, 
and find that there is a linear relation 
between the probability to be in a subregion 
and the autocorrelation time for the corresponding subsample. 
We numerically confirm this scaling law for a chiral random matrix model.
%%%%%%%%%%%%%%%%%%%%%%%%%%%%%%%%%%%%%%% 
\end{abstract}
%%%%%%%%%%%%%%%%%%%%%%%%%%%%%%%%%%%%%%% 
\end{titlepage}

\pagestyle{empty}
\pagestyle{plain}

\tableofcontents
\setcounter{footnote}{0}

%%%%%%%%%%%%%%%%%%%%%%%%%%%%%%%%%%%%%%%
%%%%%%%%%%%%%%%%%%%%%%%%%%%%%%%%%%%%%%%
\section{Introduction}
\label{sec:introduction}
%%%%%%%%%%%%%%%%%%%%%%%%%%%%%%%%%%%%%%%
%%%%%%%%%%%%%%%%%%%%%%%%%%%%%%%%%%%%%%%

The numerical sign problem has prevented us 
from the first-principles analysis of various important systems, 
such as quantum chromodynamics (QCD) at finite density \cite{Aarts:2015tyj}, 
quantum Monte Carlo calculations 
of quantum statistical systems \cite{Pollet:2012},  
and the real-time dynamics of quantum fields. 

Among various approaches to the sign problem, 
some utilize the complexification of dynamical variables. 
For example, in the complex Langevin method 
\cite{Parisi:1983cs,Klauder:1983sp,Aarts:2009dg,Nishimura:2015pba},
one considers the Langevin equation 
in the complexified configuration space. 
In the path optimization method 
\cite{Mori:2017pne,Mori:2017nwj,Alexandru:2018fqp,Bursa:2018ykf},
with the aid of machine learning 
one looks for an optimized integration surface 
for which the average phase factor is maximal. 
In the Lefschetz thimble method 
\cite{Witten:2010cx,Cristoforetti:2012su,Cristoforetti:2013wha,
Fujii:2013sra,Fujii:2015bua,Fujii:2015vha,
Alexandru:2015xva,Alexandru:2015sua,Fukuma:2017fjq,
Alexandru:2017oyw,Fukuma:2019wbv,Fukuma:2019uot,Fukuma:2020fez}, 
one deforms the integration surface 
according to the antiholomorphic gradient flow. 
The deformed surface asymptotes to a union of Lefschetz thimbles, 
each of which gives a constant value to the imaginary part of the action 
and thus is free from the sign problem. 
Although there can appear the ergodicity problem 
due to the existence of infinitely high potential barriers between thimbles, 
this ergodicity problem can be diminished 
by tempering the system with the flow time 
\cite{Fukuma:2017fjq}. 
This {\em tempered Lefschetz thimble method} (TLTM) 
thus solves the sign and ergodicity problems simultaneously. 
Moreover, 
the computational cost of TLTM has recently been reduced significantly 
by developing the {\em worldvolume tempered Lefschetz thimble method} (WV-TLTM) 
\cite{Fukuma:2020fez}, 
which we are going to review now. 

Let $\bbR^N=\{x\}$ be the configuration space 
and $S(x)$ the action (allowed to be complex-valued). 
Our main interest is to numerically estimate 
the expectation values of observables $\mathcal{O}(x)$, 
\begin{align}
  \langle \mathcal{O} \rangle
  \equiv \frac{
  \int_{\bbR^N} dx\,e^{-S(x)}\,\mathcal{O}(x)
  }{
  \int_{\bbR^N} dx\,e^{-S(x)}
  }.
\end{align}
Under the assumption that 
$e^{-S(z)}$ and $e^{-S(z)}\,\mathcal{O}(z)$ 
are entire functions of $z\in\bbC^N$, 
Cauchy's theorem allows us to continuously deform the integration surface 
without changing the value of integral. 
By expressing the deformation as a flow $z_t(x)$ $(t\geq0)$ with $z_0(x)=x$, 
the deformed integration surface at flow time $t$ 
can be written as $\Sigma_t=\{z_t(x)\,|\,x\in\bbR^N\}$, 
and thus we have the equality
\begin{align}
  \langle \mathcal{O} \rangle
  = \frac{
  \int_{\Sigma_t} dz_t\,e^{-S(z_t)}\,\mathcal{O}(z_t)
  }{
  \int_{\Sigma_t} dz_t\,e^{-S(z_t)}
  }. 
\label{eq:z_t}
\end{align}
Since the numerator and the denominator are both independent of $t$, 
we can integrate each of them over an arbitrary interval $[T_0,T_1]$ 
with an arbitrary function $W(t)$. 
This rewrites \eqref{eq:z_t} 
to the integration over the {\em worldvolume} 
$\mathcal{R} \equiv \bigcup_{T_0\leq t \leq T_1} \Sigma_t$:
 \begin{align}
  \langle \mathcal{O} \rangle
  = \frac{
  \int_{T_0}^{T_1} dt\, e^{-W(t)} \int_{\Sigma_t} dz_t\,
  e^{-S(z_t)}\,\mathcal{O}(z_t)
  }{
  \int_{T_0}^{T_1} dt\, e^{-W(t)} \int_{\Sigma_t} dz_t\,e^{-S(z_t)}
  }
  = \frac{
  \int_{\mathcal{R}} Dz \, K(z) \, e^{-W(t(z))} \,
  e^{-S(z)}\,\mathcal{O}(z)
  }{
  \int_{\mathcal{R}} Dz \, K(z) \, e^{-W(t(z))} \,e^{-S(z)}
  },
\label{eq:estimation_formula}
\end{align}
where $Dz$ is the induced volume element on $\mathcal{R}$, 
and $K(z)$ is the Jacobian: $dt\,dz_t = K(z)\,Dz$. 
The weight factor $e^{-W(t)}$ is chosen 
so that the probability for a configuration to appear at time $t$ 
is (almost) independent of $t$. 
This setting is especially necessary 
when the whole range of $t$ is relevant to simulations, 
as in the WV-TLTM. 
We further rewrite \eqref{eq:estimation_formula} 
to the ratio of reweighted averages: 
\begin{align}
  \langle \mathcal{O} \rangle
  = \frac{\int_\mathcal{R} Dz\,
  e^{-V(z)}\,A(z)\,\mathcal{O}(z)}
  {\int_\mathcal{R} Dz\,e^{-V(z)}\,A(z)}
  = \frac{\langle A(z)\,\mathcal{O}(z) \rangle_\mathcal{R}}
  {\langle A(z) \rangle_\mathcal{R}}. 
\end{align}
Here, the reweighted average  
\begin{align}
  \langle \mathcal{O}(z) \rangle_\mathcal{R} \equiv 
  \frac{\int_\mathcal{R} Dz\,e^{-V(z)}\,\mathcal{O}(z)}
  {\int_\mathcal{R} Dz\,e^{-V(z)}}
\label{expec_R}
\end{align}
is defined for the weight 
$e^{-V(z)} \equiv e^{-{\rm Re}\,S(z)-W(t(z))}$,
and $A(z)$ is the reweighting factor 
\begin{align}
  A(z) \equiv 
  \frac{ e^{-S(z)-W(t(z))} dt\, dz_t }{ e^{-V(z)} Dz } 
  = e^{-i\,{\rm Im}\,S(z)} K(z), 
\label{reweighting_factor}
\end{align}
whose explicit form can be found in Ref.~\cite{Fukuma:2020fez}. 

The reweighted average \eqref{expec_R} is estimated 
by the average over a sample 
generated  by the hybrid Monte Carlo (HMC) algorithm 
with the potential $V(z)$. 
We will generally call a HMC algorithm 
on an accumulation of integration surfaces 
the {\em worldvolume hybrid Monte Carlo algorithm} 
(the WV-HMC algorithm),
which includes the WV-TLTM.

\begin{figure}[ht]
  \centering
  \includegraphics[width=70mm]{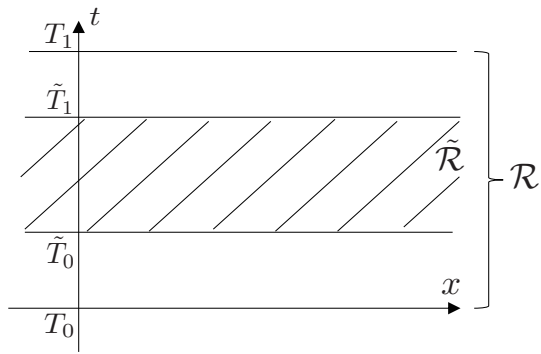} 
  \caption{
    \label{fig:worldvolume}
    The worldvolume $\mathcal{R}$ and a subregion $\tilde{\mathcal{R}}$.
  }
\end{figure}

In the case of the WV-TLTM, 
the interval $[T_0,T_1]$ should include the region with large $t$ 
so as to solve the sign problem, 
and at the same time include the region with small $t$ 
so as to reduce the ergodicity problem. 
However, the small-$t$ region may be contaminated by the sign problem, 
and the large-$t$ region by the ergodicity problem. 
Thus, in order to reduce the contributions 
from these potentially contaminated regions, 
it was proposed in Ref.~\cite{Fukuma:2020fez}
to estimate the observables 
from a subsample in a subregion with $[\tilde{T}_0,\tilde{T}_1]$ 
($\subset [T_0, T_1]$) 
(see Fig.~\ref{fig:worldvolume}), 
where $\tilde T_0$ and $\tilde T_1$ are chosen 
such that the estimated values vary only slightly 
against the changes of $\tilde T_0$ and $\tilde T_1$.% 
\footnote{%-----
  $\tilde T_{0,1}$ were written as $\hat T_{0,1}$ 
  in Ref.~\cite{Fukuma:2020fez}.
} %-------------

In Ref.~\cite{Fukuma:2020fez} 
the standard error analysis was employed 
as if the subsample itself is generated as a Markov chain 
with ergodicity and detailed balance. 
However, this is not obvious and needs a justification. 
In this paper, 
we prove that, 
if consecutive configurations in the worldvolume 
are generated as a Markov chain with ergodicity and detailed balance, 
then the subset consisting of the configurations belonging to a subregion 
can also be regarded as a Markov chain 
with ergodicity and detailed balance intact. 
We particularly investigate the integrated autocorrelation times 
for the Markov chains in various subregions, 
and find that there is a linear relation 
between the probability to be in a subregion 
and the integrated autocorrelation time for the corresponding subsample. 
We numerically confirm this scaling law for a chiral random matrix model 
(the Stephanov model \cite{Stephanov:1996ki,Halasz:1998qr}).
 
This paper is organized as follows. 
In section~\ref{sec:stochastic_subregion_autocorrelation}, 
% we first prove that the stochastic process in a subregion is a Markov chain. 
we first prove that 
the subset consisting of the configurations in a subregion 
is a Markov chain. 
We then argue that 
there should be a linear relation 
between the probability to be in a subregion 
and the integrated autocorrelation time for the corresponding subsample. 
Section~\ref{sec:confirmation} demonstrates this scaling 
by explicit numerical calculations for the Stephanov model.
Section~\ref{sec:conclusion_outlook} is devoted to conclusion and outlook. 
In appendix~\ref{sec:jackknife} 
we explain our Jackknife method 
for estimating the statistical errors of the integrated autocorrelation times.

%%%%%%%%%%%%%%%%%%%%%%%%%%%%%%%%%%%%%%%%%%%%%%%%%%%%%%%
%%%%%%%%%%%%%%%%%%%%%%%%%%%%%%%%%%%%%%%%%%%%%%%%%%%%%%%
\section{Stochastic process in a subregion 
and the integrated autocorrelation time}
\label{sec:stochastic_subregion_autocorrelation}
%%%%%%%%%%%%%%%%%%%%%%%%%%%%%%%%%%%%%%%%%%%%%%%%%%%%%%%
%%%%%%%%%%%%%%%%%%%%%%%%%%%%%%%%%%%%%%%%%%%%%%%%%%%%%%%

Let $\mathcal{R}=\{z\}$ be the full configuration space 
(the worldvolume in the case of the WV-TLTM). 
Suppose that we are given a Markov chain 
$\{z_k\}$ $(k=0,1,2,\ldots)$ in $\mathcal{R}$ 
with the transition probability $P(z_{k+1}|z_k)$ 
that satisfies ergodicity 
as well as the detailed balance condition 
with respect to the unique equilibrium distribution $p_{\rm eq}(z)$: 
\begin{align}
  P(z'|z)\, p_{\rm eq}(z) = P(z|z')\, p_{\rm eq}(z').
\end{align}
Only in this section, 
we write $\int_{\mathcal{R}} dz\,p_{\rm eq}(z)\,\mathcal{O}(z)$ 
simply by $\langle \mathcal{O}\rangle$ 
(instead of $\langle \mathcal{O}\rangle_\mathcal{R}$).

%%%%%%%%%%%%%%%%%%%%%%%%%%%%%%%%%%%%%%%%%%%%%%%%%%%%%%%
\subsection{Stochastic process in a subregion}
\label{sec:stochastic_subregion}
%%%%%%%%%%%%%%%%%%%%%%%%%%%%%%%%%%%%%%%%%%%%%%%%%%%%%%%

We now look at a subregion $\tilde{\mathcal{R}}$ in $\mathcal{R}$, 
whose complement we denote by 
$\tilde{\mathcal{R}}^c \equiv \mathcal{R} 
\backslash \tilde{\mathcal{R}}$.% 
\footnote{%-----
  In the WV-TLTM, $\mathcal{R}=\bigcup_{t=T_0}^{T_1}\Sigma_t$ 
  and $\tilde{\mathcal{R}}=\bigcup_{t=\tilde T_0}^{\tilde T_1}\Sigma_t$ 
  with $T_0\leq \tilde T_0 < \tilde T_1 \leq T_1$. 
} %-------------  
Then, from the Markov chain 
$\{z_k\}$ $(k=0,1,2,\ldots)$ in $\mathcal{R}$, 
we can extract a subsequence $\{\tilde z_\ell\}$ $(\ell=0,1,2,\ldots)$ 
that consists of the configurations belonging to $\tilde{\mathcal{R}}$. 
We first notice that 
this sequence is a Markov chain 
with the following transition probability 
from $\tilde z\in\tilde{\mathcal{R}}$ 
to $\tilde z'\in\tilde{\mathcal{R}}$: 
\begin{align}
  \tilde{P}(\tilde {z}'|\tilde{z}) = P(\tilde{z}'|\tilde{z}) 
  & + \int_{\tilde{\mathcal{R}}^c} dw \, 
  P(\tilde{z}'|w)P(w|\tilde{z}) 
\nonumber
\\
  & + \int_{\tilde{\mathcal{R}}^c} dw_2 \, dw_1 \, 
  P(\tilde{z}'|w_2)P(w_2|w_1)P(w_1|\tilde{z}) 
\nonumber
\\
  & + \cdots.
\label{eq:hatP}
\end{align}
Since $P(z'|z)$ is ergodic by assumption, 
and thus since a configuration which has left $\tilde{\mathcal{R}}$ 
will eventually reenter $\tilde{\mathcal{R}}$ 
at a finite number of steps, 
$\tilde{P}(\tilde z'|\tilde z)$ is ergodic 
and satisfies the probability conservation:
\begin{align}
  \int_{\tilde{\mathcal{R}}} d\tilde{z}'\, 
  \tilde{P}(\tilde{z}'|\tilde{z}) 
  = 1 \quad (\forall \tilde{z} \in \tilde{\mathcal{R}}). 
\end{align}
Furthermore, using the expression \eqref{eq:hatP}, 
one can easily show that 
$\tilde{P}(\tilde z'|\tilde z)$ satisfies the following equality:
\begin{align}
  \tilde P(\tilde{z}'|\tilde z)\,p_{\rm eq}(\tilde z) 
  = \tilde P(\tilde z|\tilde z')\,p_{\rm eq}(\tilde z') \quad 
  (\tilde z', \tilde z\in\tilde{\mathcal{R}}), 
\end{align}
from which we find that 
the unique equilibrium distribution $\tilde{p}_{\rm eq}(\tilde z)$ 
for $\tilde P(\tilde z'|\tilde z)$ 
is given by 
\begin{align}
  \tilde{p}_{\rm eq}(\tilde z) 
  = \frac{p_{\rm eq}(\tilde z)}
  {\int_{\tilde{\mathcal{R}}} d\tilde{z}'\, p_{\rm eq}(\tilde{z}')}. 
\end{align}
We thus have proved that 
the estimation of the expectation value 
$\int_{\tilde{\mathcal{R}}} d\tilde z\,
\tilde p_{\rm eq}(\tilde z)\,\mathcal{O}(\tilde z)$
with the subsample from a subregion $\tilde{\mathcal{R}}$ 
can be statistically analyzed 
as if it is a Markov chain.

%%%%%%%%%%%%%%%%%%%%%%%%%%%%%%%%%%%%%%%%%%%%%%%%%%%%%%%
\subsection{Integrated autocorrelation time for a subchain}
\label{sec:tauint}
%%%%%%%%%%%%%%%%%%%%%%%%%%%%%%%%%%%%%%%%%%%%%%%%%%%%%%%

Let $\{z_k\}$ $(k=0,1,2,\ldots)$ 
again be a Markov chain in $\mathcal{R}$ 
with the transition probability $P(z'|z)$. 
Denoting $\mathcal{O}(z_k)$ by $\mathcal{O}_k$, 
we define the integrated autocorrelation time of $\mathcal{O}$ by%
\footnote{ %-----
  This normalization gives the effective sample size 
  as $N^{\rm eff}_{\rm conf} = N_{\rm conf}/\tau_{\rm int}$ 
  (see, e.g., Ref.~\cite{Madras:1988ei}). 
} %-------------- 
\begin{align}
  \tau_{\rm int}(\mathcal{O}) \equiv 1 + 2\sum_{k=1}^{\infty} 
  \frac{C_k(\mathcal{O})}{C_0(\mathcal{O})}, 
\label{eq:def_tauint}
\end{align}
where 
$C_k(\mathcal{O})\equiv\langle \mathcal{O}_0\mathcal{O}_k\rangle_c \equiv
\langle (\mathcal{O}_0-\langle \mathcal{O} \rangle )
( \mathcal{O}_k-\langle \mathcal{O} \rangle ) \rangle $ 
is the autocorrelation.
Similarly, we define the integrated autocorrelation time 
for the subchain $\{\tilde z_\ell\}$ $(\ell=0,1,2,\ldots)$ 
in $\tilde{\mathcal{R}}$, 
and denote it by $\tilde\tau_{\rm int}(\mathcal{O})$.
Note that $\tilde{\tau}_{\rm int}(\mathcal{O})$ 
is generically smaller than $\tau_{\rm int}(\mathcal{O})$, 
% because one-step transition with $\tilde P$ generically corresponds 
% to transitions of multiple steps with $P$. 
because one-step transition with $\tilde P$ 
% generically corresponds 
can correspond to transitions of multiple steps with $P$. 

The ratio $\tilde\tau_{\rm int}(\mathcal{O})/\tau_{\rm int}(\mathcal{O})$ 
can be evaluated as follows, 
when both the numerator and the denominator are not too small. 
We first write by $\epsilon$ and $\tilde{\epsilon}$, respectively, 
the average Monte Carlo times evolved in one-step transition 
with $P$ and $\tilde P$.%
\footnote{ %-----
  This $\epsilon$ corresponds to the Langevin time for Langevin algorithms 
  and to the molecular dynamics time 
  multiplied by the average acceptance rate for HMC algorithms. 
} %-------------- 
We then note that $\tilde\epsilon$ can be written with $\epsilon$ 
by using the probability $p$ for a configuration in $\tilde{\mathcal{R}}$ 
to stay in $\tilde{\mathcal{R}}$ at the next step 
and the probability $q$ for a configuration in $\tilde{\mathcal{R}}^c$ 
to stay in $\tilde{\mathcal{R}}^c$ as well:
\begin{align}
  \tilde \epsilon 
  & = p\, \epsilon + (1-p) (1-q)\, 2\epsilon  
  + (1-p) q (1-q)\, 3\epsilon + \cdots 
\nonumber 
\\
  & = \frac{2-p-q}{1-q} \,\epsilon 
\nonumber 
\\
  \Bigl(&= \frac{(1-p)+(1-q)}{1-q}\,\epsilon 
   \geq \epsilon \Bigr). 
\label{eq:epsilon_scaling}
\end{align}
Furthermore, since autocorrelations should be the same 
at large Monte Carlo time scales, 
we have the equality  
\begin{align}
  \tau_{\rm int}(\mathcal{O})\, \epsilon 
  = \tilde{\tau}_{\rm int}(\mathcal{O})\, \tilde\epsilon. 
\label{eq:tauint}
\end{align}
Note that this renormalization-group-like argument holds 
only when $\tau_{\rm int}(\mathcal{O})$ 
and $\tilde\tau_{\rm int}(\mathcal{O})$ are not too small. 
Combining Eq.~\eqref{eq:epsilon_scaling} and Eq.~\eqref{eq:tauint}, 
we obtain the desired result:
\begin{align}
  \frac{\tilde{\tau}_{\rm int}(\mathcal{O})}{\tau_{\rm int}(\mathcal{O})} 
  = \frac{\epsilon}{\tilde{\epsilon}} 
  = \frac{1-q}{2-p-q} \leq 1. 
\label{eq:tauint_ratio}
\end{align}

%%%%%%%%%%%%%%%%%%%%%%%%%%%%%%%%%%%%%%%%%%%%%%%%%%%%%%%
\subsection{Scaling law for the integrated autocorrelation times}
\label{sec:scaling_law}
%%%%%%%%%%%%%%%%%%%%%%%%%%%%%%%%%%%%%%%%%%%%%%%%%%%%%%%

We now apply the preceding arguments 
to the case where the configuration space is the worldvolume of the WV-TLTM. 
We argue that there must be a linear relation 
between the probability to be in a subregion $\tilde{\mathcal{R}}$ 
and the integrated autocorrelation time $\tilde\tau_{\rm int}$ 
for the corresponding subchain. 
This claim will be confirmed numerically in the next section. 

To simplify discussions, 
we assume that the integrated autocorrelation time for the flow time $t$ 
[i.e., $\tau_{\rm int}(\mathcal{O})$ with $\mathcal{O}(z)=t(z)$] 
is sufficiently small, $\tau_{\rm int}(t) \simeq 1$. 
This can be easily realized, if necessary, 
by removing consecutive configurations 
from the chain at appropriate intervals. 
The smallness of $\tau_{\rm int}(t)$ means that 
the probability $p$ simply expresses the probability 
for a configuration to be in $\tilde{\mathcal{R}}$. 
We now recall that the distribution of $t$ is uniform in equilibrium 
for the WV-TLTM
[see the discussion below Eq.~\eqref{eq:estimation_formula}]. 
Thus, $p$ is given by 
\begin{align}
  p = \frac{ \tilde{T}_1 - \tilde{T}_0 }{ T_1 - T_0 }.
\label{eq:scaling}
\end{align}
A similar statement holds for the probability $q$ 
which now expresses the probability 
for a configuration to be in $\tilde{\mathcal{R}}^c$, 
and thus we obtain the relation $p+q=1$. 
Then, Eq.~\eqref{eq:epsilon_scaling} leads to the relation  
$\tilde\epsilon = \epsilon/p$, 
and thus, combined with Eq.~\eqref{eq:scaling}, 
we obtain the following scaling law 
for the integrated autocorrelation times:
\begin{align}
  p = \frac{ \tilde{T}_1 - \tilde{T}_0 }{ T_1 - T_0 }
  = \frac{\tilde{\tau}_{\rm int}(\mathcal{O})}{\tau_{\rm int}(\mathcal{O})}.
\label{eq:simplified_scaling}
\end{align}

We now consider the numerical estimation of $\langle\mathcal{O}\rangle$ 
using a subsample belonging to the interval $[\tilde{T}_0, \tilde{T}_1]$. 
Since Cauchy's theorem ensures $\langle \mathcal{O} \rangle$ 
to be independent of the choice of $[\tilde{T}_0, \tilde{T}_1]$, 
the estimate should not vary largely 
against the changes around an appropriately chosen pair 
$[\tilde{T}_0, \tilde{T}_1]$. 
Furthermore, 
the statistical error $\delta \langle \mathcal{O} \rangle$ 
also hardly depends on the choice of $[\tilde{T}_0, \tilde{T}_1]$. 
To see this, let us write the number of configurations 
in the interval $[\tilde{T}_0, \tilde{T}_1]$ 
as $N_{\rm conf}( \tilde{T}_0, \tilde{T}_1 )$. 
Since the distribution of $t$ is almost uniform, 
the ratio 
$N_{\rm conf}( \tilde{T}_0, \tilde{T}_1 ) / N_{\rm conf}( T_0, T_1 )$ 
almost equals $p$, 
and thus we obtain the relation 
\begin{align}
  \frac{N_{\rm conf}(\tilde{T}_0,\tilde{T}_1)}{N_{\rm conf}(T_0,T_1)} 
  \simeq \frac{\tilde{\tau}_{\rm int}(\mathcal{O})}
  {\tau_{\rm int}(\mathcal{O})}. 
\end{align}
This in turn means that the effective number of configurations 
in a subsample 
does not depend on the choice of the interval $[\tilde{T}_0,\tilde{T}_1]$, 
because 
\begin{align}
  N_{\rm conf}^{\rm eff} (\mathcal{O};\tilde{T}_0,\tilde{T}_1)
  \equiv \frac{N_{\rm conf}(\tilde{T}_0,\tilde{T}_1)}
  {\tilde{\tau}_{\rm int}(\mathcal{O})} 
  \simeq \frac{N_{\rm conf}(T_0,T_1)}{\tau_{\rm int}(\mathcal{O})}
  \quad (\forall\,\tilde{T}_0,\,\tilde{T}_1). 
\label{eq:effective_sample_size}
\end{align}
When $N_{\rm conf}$ is sufficiently large (as we always assume), 
the statistical error is given by the formula 
$\delta \langle \mathcal{O} \rangle 
=  \sigma_{\mathcal{O}} 
   / \sqrt{ N_{\rm conf}^{\rm eff}(\mathcal O; \tilde T_0, \tilde T_1) }$ 
with a constant $\sigma_{\mathcal{O}}$ 
($=\sqrt{\langle \mathcal{O}^2 \rangle_c}$). 
Since $N_{\rm conf}^{\rm eff}(\mathcal O; \tilde T_0, \tilde T_1)$ 
is almost independent of the choice 
of $[\tilde{T}_0,\tilde{T}_1]$, 
so is $\delta \langle \mathcal{O} \rangle$.

%%%%%%%%%%%%%%%%%%%%%%%%%%%%%%%%%%%%%%%%%%%%%%%%%%%%%%%
%%%%%%%%%%%%%%%%%%%%%%%%%%%%%%%%%%%%%%%%%%%%%%%%%%%%%%%
\section{Numerical confirmation of the scaling law}
\label{sec:confirmation}
%%%%%%%%%%%%%%%%%%%%%%%%%%%%%%%%%%%%%%%%%%%%%%%%%%%%%%%
%%%%%%%%%%%%%%%%%%%%%%%%%%%%%%%%%%%%%%%%%%%%%%%%%%%%%%%

We numerically confirm the scaling law \eqref{eq:simplified_scaling} 
for the WV-TLTM~\cite{Fukuma:2020fez} 
applied to a chiral random matrix model 
(the Stephanov model \cite{Stephanov:1996ki,Halasz:1998qr}).

%%%%%%%%%%%%%%%%%%%%%%%%%%%%%%%%%%%%%%%%%%%%%%%%%%%%%%%
\subsection{Setup}
%%%%%%%%%%%%%%%%%%%%%%%%%%%%%%%%%%%%%%%%%%%%%%%%%%%%%%%

The partition function of the Stephanov model 
for $N_f$ quarks with equal mass $m$ is given by 
\begin{align}
  Z_n^{N_f} = e^{n\mu^2}\int d^2 X\,e^{-S(X,X^\dagger)}
  \equiv e^{n\mu^2}\int d^2 X\,e^{-n\,{\rm tr}\,X^\dagger X}\,
  {\det}^{N_f}(D+m), 
\label{stephanov_partition}
\end{align}
where $X=(X_{ij}=x_{ij}+i\,y_{ij})$ is an $n\times n$ complex matrix. 
The $2n \times 2n$ matrix $D$ expresses the Dirac operator 
in the chiral representation,
\begin{align}
  & D \equiv \left(
    \begin{array}{cc}
      0 & i\,(X+C)\\
      i\,(X^\dagger+C) & 0
    \end{array}\right), \\
  & i\,C \equiv \left(
  \begin{array}{cc}
    (\mu+i\tau)\,1_{n/2} & 0\\
    0 & (\mu-i\tau)\,1_{n/2}
  \end{array}\right),
\end{align}
where $\mu$ and $\tau$ represent 
the chemical potential and the temperature, respectively.
The chiral condensate and the number density 
are defined, respectively, by 
\begin{align}
  \langle \bar\psi \psi \rangle
  &\equiv \frac{1}{2n}\,\frac{\partial}{\partial m}\,
  \ln Z_n^{N_f},
  \quad 
  \langle \psi^\dagger \psi \rangle
  \equiv \frac{1}{2n}\,\frac{\partial}{\partial \mu}\,
  \ln Z_n^{N_f}.
\end{align} 
We will set the parameters 
to $N_f=1$, $n=2$, $\mu=0.6$, $\tau=0$, $m=0.004$. 

We generate a sample $\{z_j\}$ $(j=1,\ldots,N_{\rm conf})$ 
with the HMC algorithm using the potential $V(z)$, 
and estimate the reweighted average 
$\langle \mathcal{O}(z) \rangle_\mathcal{R}$ 
[see Eq.~\eqref{expec_R}]
by the sample average 
\begin{align}
  \overline{\mathcal{O}} \equiv \frac{1}{N_{\rm conf}}
  \sum_{j=1}^{N_{\rm conf}}\mathcal{O}(z_j). 
\end{align}
The estimator of the integrated autocorrelation time 
$\tau_{\rm int}(\mathcal{O})$
is given by 
\begin{align}
  \overline\tau_{\rm int}(\mathcal{O};k_{\rm max})
  \equiv 1 + 2\sum_{k=1}^{k_{\rm max}}
  \frac{ \overline{C}_k( \mathcal{O} ) }
  { \overline{C}_0( \mathcal{O} ) }.
\end{align}
Here, $\overline{C}_k(\mathcal{O})$ 
is an estimator of the autocorrelation  
$C_k(\mathcal{O})
=\langle \mathcal{O}_0\mathcal{O}_k\rangle_{\mathcal{R},\,c}$, 
whose explicit form is given in appendix~\ref{sec:jackknife}.
We have truncated the summation at $k_{\rm max}$ 
to avoid summing up statistical fluctuations around zero 
(see, e.g., Ref.~\cite{Priestley:1981}).
The statistical error $\delta\overline\tau_{\rm int}(\mathcal{O};k_{\rm max})$ 
is estimated 
by a Jackknife method that is described in appendix~\ref{sec:jackknife}. 
Values of $k_{\rm max}$ and bin sizes used in the estimations of 
$\overline\tau_{\rm int}(\mathcal{O};k_{\rm max})$ are summarized 
in Table \ref{table:kmax_binsize}. 
\begin{table}[ht]
  \centering
  %-----------------
  \begin{small}
    \begin{tabular}{|c|c|c|c|c|c|c|c|c|c|c|c|}
      \hline
      $\mathcal{O}$& $i$& 0& 1& 2& 3& 4& 5& 6& 7& 8& 9 \\
      \hline
      $t$
                   & $k_{\rm max}$& 75& 60& 60& 60& 60& 60& 50& 50& 50& 50 \\
                   & bin size& 140& 140& 140& 140& 140& 120& 120& 100& 100& 100\\
      ${\rm Re}\,(\bar{\psi}\psi A)$
                   & $k_{\rm max}$& 900& 250& 250& 250& 250& 300& 300& 300& 400& 500 \\
                   & bin size& 1100& 800& 900& 500& 900& 500& 600& 500& 500& 400 \\
      ${\rm Re}\,(\psi^\dagger\psi A)$
                   & $k_{\rm max}$& 500& 500& 500& 500& 500& 500& 500& 500& 500& 500 \\
                   & bin size& 1000& 1300& 1200& 1300& 1100& 1100& 1100& 1100& 1100& 1300 \\
      ${\rm Re}\,A$
                   & $k_{\rm max}$& 500& 500& 500& 500& 500& 500& 500& 500& 500& 500 \\
                   & bin size& 1000& 1000& 900& 700& 700& 600& 600& 500& 500& 500 \\
      \hline
      $\mathcal{O}$& $i$& 10& 11& 12& 13& 14& 15& 16& 17& 18& 19 \\
      \hline
      $t$
                   & $k_{\rm max}$& 50& 50& 50& 50& 50& 40& 30& 25& 25& 15 \\
                   & bin size& 100& 100& 100& 100& 120& 80& 100& 120& 100& 100 \\
      ${\rm Re}\,(\bar{\psi}\psi A)$
                   & $k_{\rm max}$& 500& 400& 500& 500& 300& 250& 400& 250& 400& 300 \\
                   & bin size& 400 & 400& 300& 300& 300& 300& 400& 400& 400& 400 \\
      ${\rm Re}\,(\psi^\dagger\psi A)$
                   & $k_{\rm max}$& 500& 500& 500& 500& 500& 500& 500& 500& 250& 400 \\
                   & bin size& 1100& 1100& 1300& 1300& 1100& 1100& 1100& 1100& 1100& 1100 \\
      ${\rm Re}\,A$
                   & $k_{\rm max}$& 500& 500& 500& 500& 500& 500& 500& 250& 250& 250 \\
                   & bin size& 500& 400& 500& 400& 400& 400& 400& 500& 400& 500 \\
      \hline
    \end{tabular}
  \end{small}
  % -----------------
  \caption{Values of $k_{\rm max}$ and bin sizes.}
  \label{table:kmax_binsize}
\end{table}\noindent

In the numerical simulation with the HMC, 
we set $T_0 = 0$ and $T_1 = 0.02$.
The HMC updates are performed 
with the molecular dynamics time increment $\Delta s=0.001$ 
and the step number $n_{\rm HMC}=100$ 
with the average acceptance rate more than $0.99$. 
We employ 20 independent sets of configurations, 
each set consisting of $3 \times 10^6$ configurations in $[T_0, T_1]$.
The observables are measured at every 6 iterations of the HMC algorithm, 
so that we have 20 independent samples 
of size $N_{\rm conf} = 5 \times 10^5$. 
We fix $\tilde{T}_1 = T_1$ and vary $\tilde{T}_0$ 
as $\tilde{T}_0^{(i)} \equiv (i / 20) T_1$ ($i = 0, ..., 19$), 
for each of which an independent set of configurations is used.

%%%%%%%%%%%%%%%%%%%%%%%%%%%%%%%%%%%%%%%%%%%%%%%%%%%%%%%
\subsection{Results}
%%%%%%%%%%%%%%%%%%%%%%%%%%%%%%%%%%%%%%%%%%%%%%%%%%%%%%%

\begin{figure}[ht]
  \centering
  \includegraphics[width=80mm]{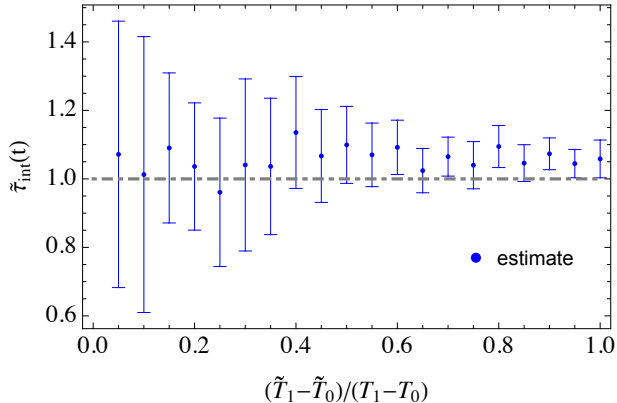} 
  \caption{
    \label{fig:tauint_t}
    Results of $\tilde{\tau}_{\rm int}(t)$. 
  }
\end{figure}

\begin{figure}[ht]
\centering
\includegraphics[width=80mm]{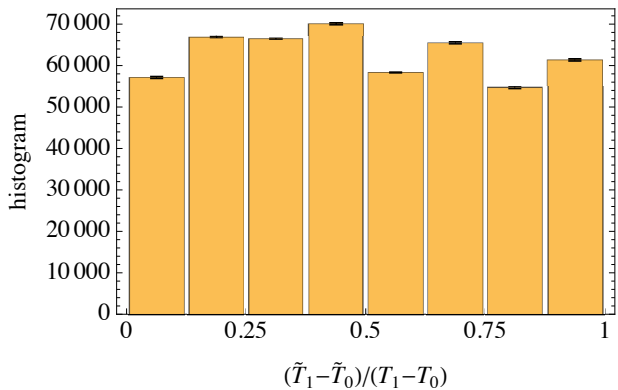} 
\caption{
\label{fig:histogram_t}
  Histogram of $t$. 
  The number of configurations in a bin 
  is the average over 20 independent samples, 
  each consisting of $N_{\rm conf} = 5 \times 10^5$.
  The function $W(t)$ is tuned 
  such that the histogram becomes almost flat for these eight bins. 
}
\end{figure}

We now demonstrate the scaling law \eqref{eq:simplified_scaling} 
from explicit numerical calculations. 
Recall that the argument for the scaling is 
based on the smallness of the integrated autocorrelation time of $t$ 
($\tau_{\rm int}(t)\simeq 1$) 
and the uniformity of the distribution of $t$. 
Figure~\ref{fig:tauint_t} shows 
that the condition $\tilde{\tau}_{\rm int}(t) \simeq 1$ is satisfied 
for all $p = (\tilde{T}_1 - \tilde{T}_0 )/( T_1 - T_0 )$. 
\noindent
Figure~\ref{fig:histogram_t} 
is the histogram of $p=(\tilde{T}_1 - \tilde{T}_0 )/( T_1 - T_0 )$, 
which is almost flat as required. 
\noindent
%
% In the simulations, 
% we have tuned the functional form of $W(t)$ 
This is realized by tuning the functional form of $W(t)$ 
as in Ref.~\cite{Fukuma:2020fez}.

\begin{figure}[ht]
  \centering
  \includegraphics[width=70mm]{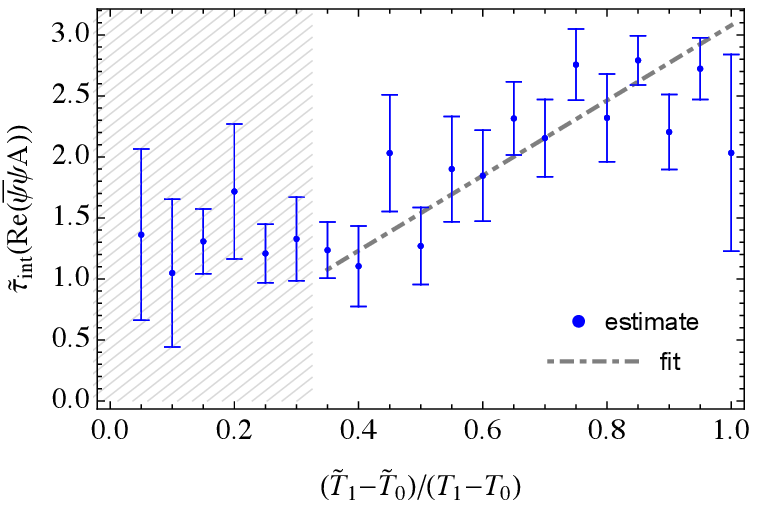} 
  \vspace{6mm}
  \includegraphics[width=70mm]{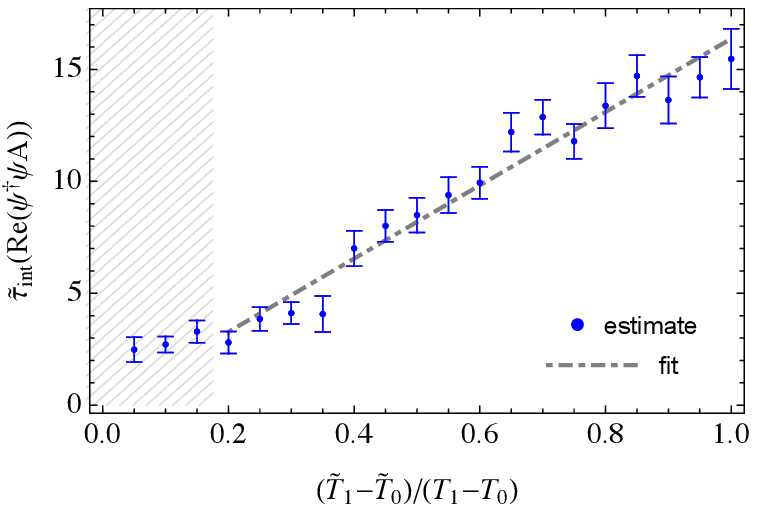} \\
  \includegraphics[width=70mm]{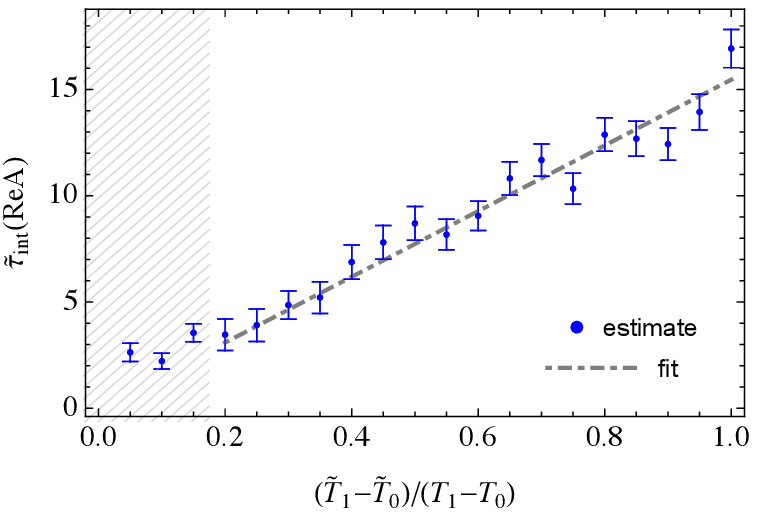} 
  \caption{
    \label{fig:tauint}
    $\tilde{\tau}_{\rm int}(\mathcal{O})$ against 
    $(\tilde{T}_1 - \tilde{T}_0 )/( T_1 - T_0 )$. 
    Top left: $\mathcal{O} = {\rm Re}\, (\bar{\psi}\psi A )$. 
    Top right: $\mathcal{O} = {\rm Re}\, (\psi^\dagger\psi A)$.
    Bottom: $\mathcal{O} = {\rm Re}\, A$. 
    Data points in the shaded region give 
    $\tilde\tau_{\rm int}(\mathcal{O})=O(1)$  
    and thus are excluded from the linear fitting 
    [see the comment below Eq.~\eqref{eq:tauint}]. 
  }
\end{figure}

Figure~\ref{fig:tauint} exhibits the scaling law 
\eqref{eq:simplified_scaling} 
for three operators, 
$\mathcal{O} = {\rm Re}\, (\bar{\psi}\psi A)$, 
$\mathcal{O} = {\rm Re}\, (\psi^\dagger\psi A)$, 
and $\mathcal{O} = {\rm Re}\, A$.
We see that the scaling is satisfied
for $\mathcal{O} = {\rm Re}\, (\bar{\psi}\psi A)$ 
in the region $(\tilde{T}_1 - \tilde{T}_0 )/( T_1 - T_0 ) \geq 0.35$, 
and for $\mathcal{O} = {\rm Re}\, (\psi^\dagger\psi A)$ 
and $\mathcal{O} = {\rm Re}\, A$ 
in the region $(\tilde{T}_1 - \tilde{T}_0 )/( T_1 - T_0 ) \geq 0.20$. 
Deviations from the scaling  
at small $(\tilde{T}_1 - \tilde{T}_0 )/( T_1 - T_0 )$ 
should be due to %the violation of the condition 
$\tilde\tau_{\rm int}(\mathcal{O}) = O(1)$ 
[see the comment below Eq.~\eqref{eq:tauint}].
\noindent
We perform the $\chi^2$-fit to these data points with
\begin{align}
  \chi^2(\mathcal{O}) \equiv \sum_{i=0}^{i_{\rm max}}
  \frac
  { [\tilde{\tau}^{(i)}_{\rm int}(\mathcal{O})
  - c\, (\tilde{T}_1 - \tilde{T}_0^{(i)} )/( T_1 - T_0 ) ]^2 }
  { [\delta\tilde\tau_{\rm int}^{(i)}(\mathcal{O})]^2 }, 
\end{align}
where $\tilde\tau_{\rm int}^{(i)}(\mathcal{O})$ is the integrated autocorrelation time 
for the subsample with the interval $[\tilde T_0,\tilde T_1]=[\tilde T_0^{(i)},T_1]$. 
The fit results are the following: 
For $\mathcal{O} = {\rm Re}\, (\bar{\psi}\psi A)$ with $i_{\rm max} = 13$, 
$c = 3.1$ and $\chi^2/{\rm DOF} = 1.0$. 
For $\mathcal{O} = {\rm Re}\, (\psi^\dagger\psi A)$ with $i_{\rm max} = 16$, 
$c = 16.4$ and $\chi^2/{\rm DOF} = 1.2$. 
For $\mathcal{O} = {\rm Re}\, A$ with  $i_{\rm max} = 16$, 
$c = 15.5$ and $\chi^2/{\rm DOF} = 1.1$. 
These data support the scaling law~\eqref{eq:simplified_scaling}.

\begin{figure}[ht]
  \centering
  \includegraphics[width=65mm]{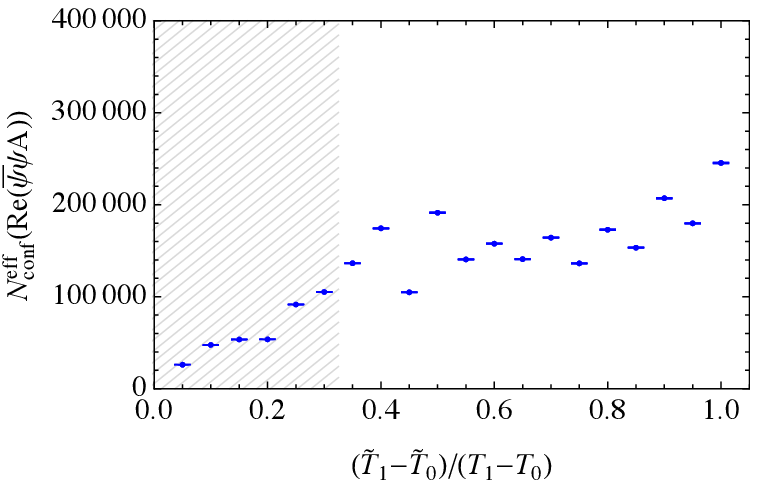} 
  \vspace{5mm}
  \includegraphics[width=65mm]{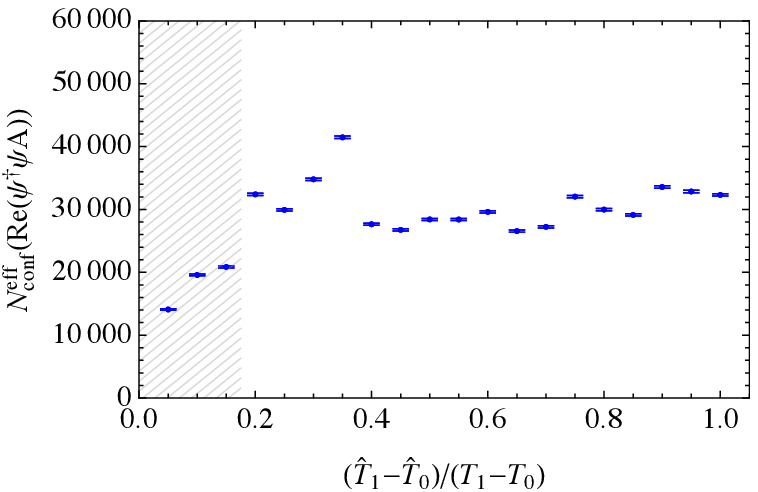} \\
  \vspace{5mm}
  \includegraphics[width=65mm]{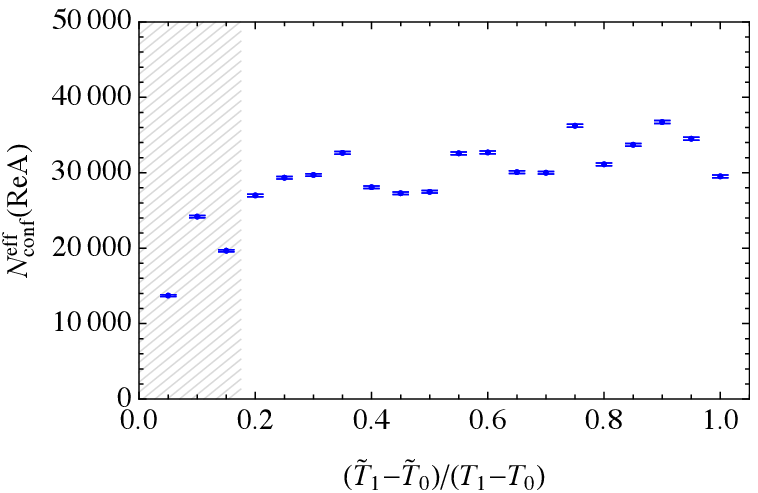} 
  \caption{
    \label{fig:Nconfeff}
    The values of 
    $N_{\rm conf}^{\rm eff}(\mathcal{O};\tilde T_0,\tilde T_1)$ 
    [see Eq.~\eqref{eq:effective_sample_size}]. 
    Top left: $\mathcal{O}={\rm Re}\, (\bar\psi \psi A)$. 
    Top right: $\mathcal{O}={\rm Re}\, (\psi^\dag \psi A)$. 
    Bottom: $\mathcal{O}= {\rm Re}\, A$. 
    The regions where the scaling law \eqref{eq:simplified_scaling} 
    is broken are shaded. 
  }
\end{figure}

We plot in Fig.~\ref{fig:Nconfeff} 
the values of $N_{\rm conf}^{\rm eff}(\mathcal{O};\tilde T_0,\tilde T_1)$ 
for various operators $\mathcal{O}$ [see Eq.~\eqref{eq:effective_sample_size}]. 
The statistical errors are estimated with the Jackknife method. 
We observe, as expected, that 
$N_{\rm conf}^{\rm eff} (\mathcal{O};\tilde{T}_0,\tilde{T}_1)$ 
takes almost the same values 
for each $\mathcal O$ 
in the range where we observe the scaling law. 
\noindent

\begin{figure}[ht]
  \centering
  \includegraphics[width=70mm]{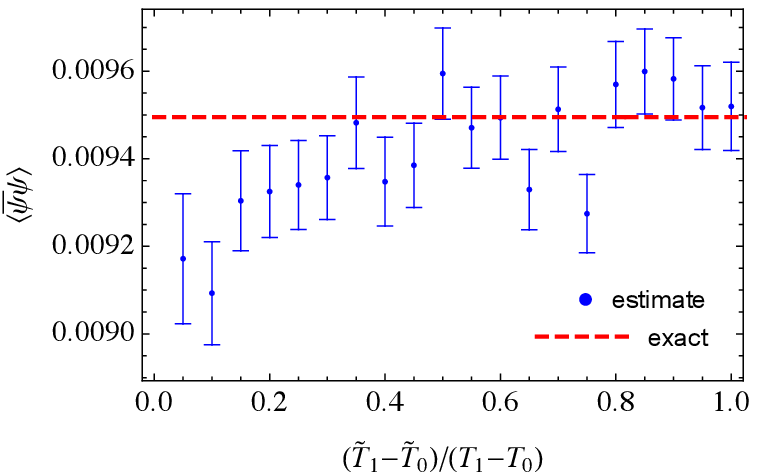} 
  \vspace{5mm}
  \includegraphics[width=70mm]{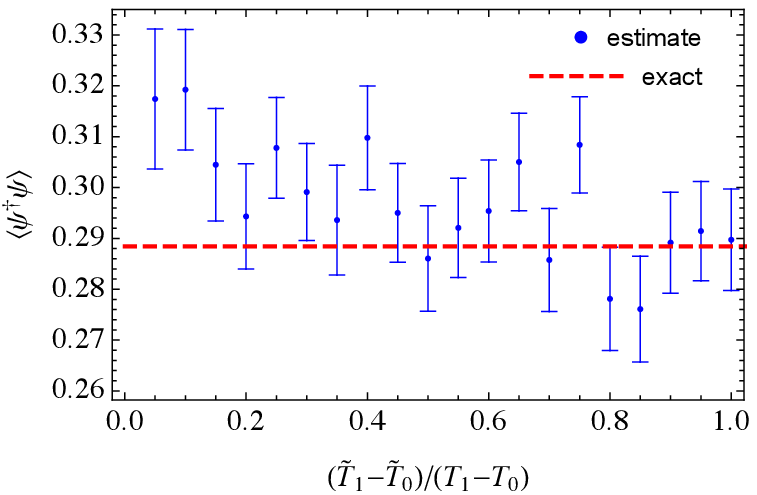} \\
  \includegraphics[width=70mm]{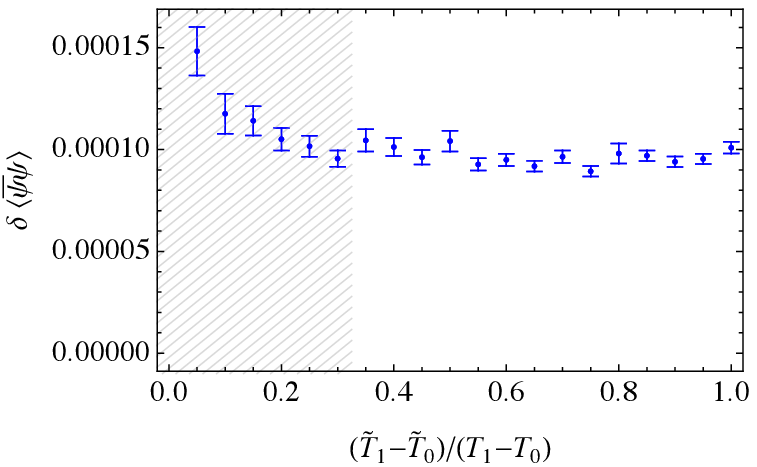} 
  \vspace{5mm}
  \includegraphics[width=70mm]{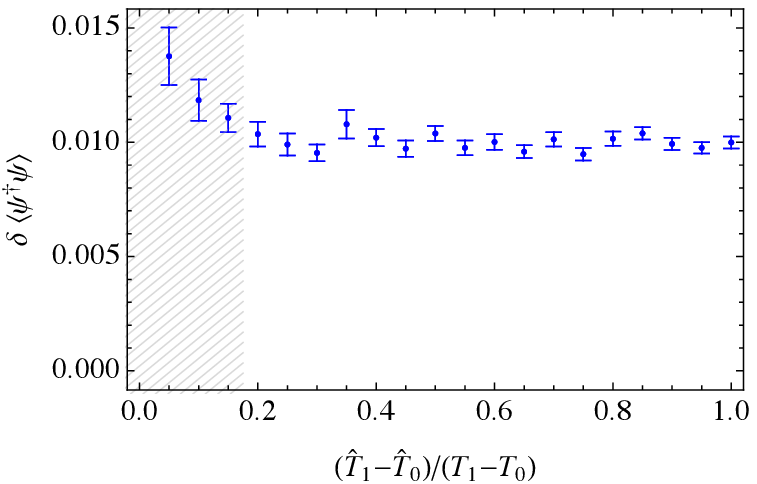} 
  \caption{
    \label{fig:estimate}
    The estimate (top) and its statistical error (bottom) 
    of the expectation value $\langle \mathcal{O} \rangle$ 
    against $(\tilde{T}_1 - \tilde{T}_0 )/( T_1 - T_0 )$. 
    Left: $\mathcal{O}=\bar\psi \psi$ (the chiral condensate). 
    Right: $\mathcal{O}=\psi^\dag \psi$ (the number density). 
    The regions where the scaling law \eqref{eq:simplified_scaling} 
    is broken in the numerator and/or the denominator 
    are shaded for the statistical errors. 
  }
\end{figure}

Finally, in Fig.~\ref{fig:estimate} 
we plot the expectation values of the chiral condensate 
and the number density, together with their statistical errors. 
\noindent
The statistical errors are estimated with the Jackknife method 
with the bin sizes fixed to 500. 
We see that 
both the means and the statistical errors take almost constant values 
in a region where $(\tilde{T}_1 - \tilde{T}_0 )/( T_1 - T_0 )$ is not small. 
The deviation of the means should be attributed 
to the complex geometry at large flow times, 
which requires larger statistics 
and better control of systematic errors 
(such as those from numerical integrations of the flow equation 
and of the Hamilton equations accompanied by the projection 
from $\bbC^N$ to $\mathcal{R}$).
The deviation of the statistical errors 
is due to the violation of the scaling law \eqref{eq:simplified_scaling} 
for either of the numerator or the denominator (or both) 
in the ratio of the reweighted averages.

%%%%%%%%%%%%%%%%%%%%%%%%%%%%%%%%%%%%%%%%%%%%%%%%%%%%%%%
%%%%%%%%%%%%%%%%%%%%%%%%%%%%%%%%%%%%%%%%%%%%%%%%%%%%%%%
\section{Summary and outlook}
\label{sec:conclusion_outlook}
%%%%%%%%%%%%%%%%%%%%%%%%%%%%%%%%%%%%%%%%%%%%%%%%%%%%%%%
%%%%%%%%%%%%%%%%%%%%%%%%%%%%%%%%%%%%%%%%%%%%%%%%%%%%%%%

In this paper, 
we have established the statistical analysis method 
for the WV-HMC algorithm, 
whose major use is intended for the WV-TLTM \cite{Fukuma:2020fez}. 
We proved that, 
if consecutive configurations in the worldvolume 
are generated as a Markov chain with ergodicity and detailed balance, 
then the subset consisting of the configurations belonging to a subregion 
can also be regarded as a Markov chain 
with ergodicity and detailed balance intact. 
We particularly investigated the integrated autocorrelation times 
for the Markov chains in various subregions, 
and found that there is a linear relation 
between the probability to be in a subregion 
and the integrated autocorrelation time for the corresponding subsample. 
We numerically confirmed this scaling law for the Stephanov model.
 
Now with this statistical analysis method at hand, 
we can safely apply the WV-TLTM to large-scale simulations 
of the systems that have serious sign problems, 
such as finite density QCD, 
strongly correlated electron systems, frustrated spin systems, 
and the real-time dynamics of quantum fields.
A study along this line is now in progress and will be reported elsewhere.

%%%%%%%%%%%%%%%%%%%%%%%%%%%%%%%%%%%%%%%%%%%%%%%%%%%%%%%
%%%%%%%%%%%%%%%%%%%%%%%%%%%%%%%%%%%%%%%%%%%%%%%%%%%%%%%
\section*{Acknowledgments}
The authors thank Issaku Kanamori, Yoshio Kikukawa and Jun Nishimura 
for useful discussions.
This work was partially supported by JSPS KAKENHI 
(Grant Numbers 18J22698, 20H01900, 21K03553) 
and by SPIRITS 
(Supporting Program for Interaction-based Initiative Team Studies) 
of Kyoto University (PI: M.F.). 
N.M.\ is supported by the Special Postdoctoral Researchers Program 
of RIKEN.
%%%%%%%%%%%%%%%%%%%%%%%%%%%%%%%%%%%%%%%%%%%%%%%%%%%%%%%
%%%%%%%%%%%%%%%%%%%%%%%%%%%%%%%%%%%%%%%%%%%%%%%%%%%%%%%

\appendix

%%%%%%%%%%%%%%%%%%%%%%%%%%%%%%%%%%%%%%%%%%%%%%%%%%%%%%%
%%%%%%%%%%%%%%%%%%%%%%%%%%%%%%%%%%%%%%%%%%%%%%%%%%%%%%%
\section{Jackknife method for the integrated autocorrelation times}
\label{sec:jackknife}
%%%%%%%%%%%%%%%%%%%%%%%%%%%%%%%%%%%%%%%%%%%%%%%%%%%%%%%
%%%%%%%%%%%%%%%%%%%%%%%%%%%%%%%%%%%%%%%%%%%%%%%%%%%%%%%

In this appendix, 
we give a Jackknife method to estimate the integrated autocorrelation times 
$\tau_{\rm int}(\mathcal{O})$.

Let $\{x_j\}$ $(j=1,\ldots,N_{\rm conf})$ be 
a set of consecutive configurations generated as a Markov chain. 
We estimate the expectation value $\langle \mathcal{O} \rangle$ 
by the sample average 
\begin{align}
  \overline{\mathcal{O}} \equiv \frac{1}{N_{\rm conf}}
  \sum_{j=1}^{N_{\rm conf}}\mathcal{O}(x_j). 
\label{eq:autocorrelation_function_1pt}
\end{align}
The estimator of the integrated autocorrelation time 
$\tau_{\rm int}(\mathcal{O})
 \equiv 1 + 2\sum_{k=1}^{\infty}
  C_k(\mathcal{O}) / C_0(\mathcal{O})$
is given by 
\begin{align}
  \bar\tau_{\rm int}(\mathcal{O};k_{\rm max})
  \equiv 1 + 2\sum_{k=1}^{k_{\rm max}}
  \frac{ \overline{C}_k( \mathcal{O} ) }
  { \overline{C}_0( \mathcal{O} ) },
\label{eq:tauint_estimate}
\end{align}
where $\overline{C}_k(\mathcal{O})$ 
is the estimator of the autocorrelation 
$C_k(\mathcal{O}) \equiv \langle \mathcal{O}_0\,\mathcal{O}_k \rangle_c$. 
The summation is truncated at $k_{\rm max}$ 
to avoid summing up statistical fluctuations around zero 
(see, e.g., Refs.~\cite{Grenander:1953,Priestley:1981}).
The value of $k_{\rm max}$ should not be set very large 
compared to $\tau_{\rm int}(\mathcal{O})$, 
otherwise contributions from statistical fluctuations around zero 
may dominate the error. 
There has been known an explicit formula 
for the statistical error $\delta \tau_{\rm int}(\mathcal{O})$ 
when $1 \ll k_{\rm max} \ll N_{\rm conf}$ 
(more precisely, as $N_{\rm conf}\to\infty$, 
$k_{\rm max}\to\infty$ and $k_{\rm max}/N_{\rm conf}\to 0$) 
\cite{Madras:1988ei} 
(see also Refs.~\cite{Grenander:1953,Priestley:1981}),% 
\footnote{ %-----
  This is given by 
  $\delta \tau_{\rm int}(\mathcal{O})
  = \tau_{\rm int}(\mathcal{O})\,\sqrt{2(2k_{\rm max} +1)/N_{\rm conf}}$ 
  for the estimator of the autocorrelation, 
  $\overline{C}_k( \mathcal{O} ) = (N_{\rm conf}-k)^{-1}\,
  \sum_{n=1}^{N_{\rm conf}-k}
  ( \mathcal{O}(x_n) - \overline{\mathcal{O}} )
  ( \mathcal{O}(x_{n+k}) - \overline{\mathcal{O}} )$ 
  \cite{Madras:1988ei}. 
} %--------------
but this may not be applicable to the case when $\tau_{\rm int}(\mathcal{O})=O(1)$, 
for which the condition $k_{\rm max}\gg 1$ cannot be met. 
Therefore, in this paper 
we adopt the Jackknife method for the estimation of 
$\delta\tau_{\rm int}(\mathcal{O})$. 

In order to apply a resampling method of Jackknife, 
we introduce a sample of multidimensional observables 
$\{X_r = (X_{r,k}^{\mathcal{O}\mathcal{O}}, 
X_{r}^{\mathcal{O}})\}$ 
($r=1,\ldots,N_{\rm conf}-k_{\rm max}$) with 
\begin{align}
  &X_{r,k}^{\mathcal{O}\mathcal{O}} \equiv
  \frac{1}{k_{\rm max} - k + 1} \sum_{i=0}^{k_{\rm max}-k} 
  \mathcal{O}(x_{r+i}) \mathcal{O}(x_{r+k+i}) 
  \quad (k=0,\ldots,k_{\rm max}), 
\\
  &X_{r}^{\mathcal{O}} \equiv
  \frac{1}{k_{\rm max} + 1} \sum_{i=0}^{k_{\rm max}} \mathcal{O}(x_{r+i}). 
\end{align}
Since 
$\langle X_{r,k}^{\mathcal{O}\mathcal{O}} \rangle
 = \langle \mathcal{O}(x_{0})\, \mathcal{O}(x_{k}) \rangle$ 
and 
$\langle X_{r}^{\mathcal{O}} \rangle = \langle \mathcal{O}(x) \rangle$, 
the autocorrelations 
$C_k(\mathcal{O})=\langle \mathcal{O}(x_0) \mathcal{O}(x_k) \rangle_c$ 
can be estimated by 
\begin{align}
  \overline{C}^{(X)}_k(\mathcal{O}) 
  \equiv \frac{1}{N_{\rm conf}-k_{\rm max}} 
  \sum_{r=1}^{N_{\rm conf}-k_{\rm max}} X_{r,k}^{\mathcal{O}\mathcal{O}}
  - \bigg( \frac{1}{N_{\rm conf}-k_{\rm max}}
  \sum_{r=1}^{N_{\rm conf}-k_{\rm max}} X_{r}^{\mathcal{O}} \bigg)^2. 
\label{eq:estimate_Jackknife}
\end{align}
Since $\tau_{\rm int}(\mathcal{O})$ is 
a function of $C_k(\mathcal{O})$, 
it can be estimated with $\overline{C}^{(X)}_k(\mathcal{O})$ as
\begin{align}
  \tau_{\rm int}(\mathcal{O}) \approx
  1 + 2\sum_{k=1}^{k_{\rm max}} 
  \frac{ \overline{C}^{(X)}_k(\mathcal{O}) }{ \overline{C}^{(X)}_0(\mathcal{O}) }. 
  \label{eq:tau_estimate_with_X}
\end{align}
The statistical error $\delta\tau_{\rm int}(\mathcal{O})$ 
can then be estimated with the standard Jackknife method. 

%%%%%%%%%%%%%%%%%%%%%%%%%%%%%%%%%%%%%%%%%%%%%%%%%%%%%%%
\baselineskip=0.9\normalbaselineskip
%%%%%%%%%%%%%%%%%%%%%%%%%%%%%%%%%%%%%%%%%%%%%%%%%%%%%%%

%%%%%%%%%%%%%%%%%%%%%%%%%%%%%%%%%%%%%%%%%%%%%%%%%%%%%%%
%%%%%%%%%%%%%%%%%%%%%%%%%%%%%%%%%%%%%%%%%%%%%%%%%%%%%%%

%%%%%%%%%%%%%%%%%%%%%%%%%%%%%%%%%%%%%%%%%%%%%%%%%%%%%%%
%%%%%%%%%%%%%%%%%%%%%%%%%%%%%%%%%%%%%%%%%%%%%%%%%%%%%%%

\end{document}